\documentclass[aps,prd,onecolumn,superscriptaddress,nofootinbib]{revtex4-2}

\usepackage{amsmath,amssymb}
\usepackage{graphicx}
\usepackage{bm}
\usepackage[colorlinks=true,citecolor=blue,linkcolor=blue,urlcolor=blue]{hyperref}
\usepackage{xcolor}

\newcommand{\eiso}{E_{k,\rm iso}}
\newcommand{\egprompt}{E_{\gamma,\rm prompt}}
\newcommand{\nism}{n_{\rm ISM}}
\newcommand{\epsB}{\epsilon_B}
\newcommand{\epse}{\epsilon_e}
\newcommand{\xie}{\xi_e}

\begin{document}

\title{Synchrotron and Inverse-Compton Signatures of Black-to-White Hole Explosions:\\
Weibel-Mediated External Shocks in the Interstellar Medium}

\author{Hoang Van Quyet}
\affiliation{Department of Physics, Hanoi Pedagogical University 2, Xuan Hoa, Phu Tho, Vietnam}
\email{hoangvanquyet@hpu2.edu.vn}

\date{\today}

\begin{abstract}
We study the non-thermal electromagnetic counterpart expected when the relativistic, near-isotropic ejecta of a black-to-white hole transition sweep up the surrounding interstellar medium (ISM). Building on a photospheric model for the prompt $\gamma$-ray emission~\cite{Villani2026}, we parametrize the kinetic energy available to drive an external shock, $\eiso=R_k\egprompt$, relative to the energy already radiated promptly, and model the resulting collisionless forward shock with a Weibel-mediated magnetic field~\cite{Weibel1959,MedvedevLoeb1999,SironiSpitkovsky2011}. We estimate the synchrotron and synchrotron self-Compton (SSC) emission from shock-accelerated electrons within a one-zone, Thomson-regime approximation, including a first-order inverse-Compton cooling correction, an order-of-magnitude Klein-Nishina suppression, synchrotron self-absorption, and an order-of-magnitude estimate of the internal $\gamma$-$\gamma$ pair opacity. For illustrative parameters ($\egprompt\sim10^{44}$ erg, $R_k=0.1$, $\Gamma_0=100$), the non-thermal afterglow's intrinsic spectrum peaks in the soft X-ray/EUV band within a fraction of a second to tens of seconds -- several orders of magnitude faster than a classical long-duration gamma-ray burst (GRB) afterglow for the adopted energy, density, and Lorentz-factor range, though the precise duration and flux depend sensitively on $R_k$, $\nism$, and the microphysical parameters $\epse,\epsB,\xie$, and Galactic photoelectric absorption can substantially suppress the observable EUV flux for a nearby source. We identify parameter regions in which an illustrative photon-count estimate exceeds a nominal detection threshold, and discuss observational features -- a fast, faint, roughly isotropic non-thermal transient following a hard-spectrum prompt flash, without an achromatic jet break -- that may help distinguish a black-to-white hole explosion from short GRBs, magnetar giant flares, and other fast transients, while flagging the assumptions (forward-shock-only dynamics, one-zone Thomson-regime radiation, an illustrative rather than instrument-specific detectability criterion) that limit the quantitative reach of the estimate.
\end{abstract}

\maketitle

\section{Introduction}
\label{sec:intro}

The possibility that black holes end their lives as white holes was proposed by Rovelli and Vidotto~\cite{RovelliVidotto2014} within Loop Quantum Gravity, following earlier work on the quantum bounce in Loop Quantum Cosmology~\cite{AshtekarPawlowskiSingh2006}. Christodoulou~et~al.~\cite{ChristodoulouEtal2016} showed that, in the external observer's frame, the lifetime of such a black hole scales as the square of its initial mass, so that lunar-mass primordial black holes (PBHs, $M\sim10^{26}$ g) have an appreciable probability of completing the transition within the current age of the Universe.

Villani~\cite{Villani2026} modeled the prompt electromagnetic signal of this transition as photospheric emission from a nearly isotropic, low-baryon-load fireball, self-consistently solving the coupled transfer equations for photons and $e^{\pm}$ pairs, and found non-thermal, quasi-thermal spectra peaking at $0.5$--$1$ MeV with sharp cutoffs at $1$--$3$ MeV, with total radiated energies $\egprompt\gtrsim10^{43}$--$10^{44}$ erg. Because no large-scale magnetic field is expected around an isolated PBH, the ejecta are not collimated into a jet; instead, a quasi-spherical shell can expand relativistically into the ISM, driving a collisionless forward shock. Ref.~\cite{Villani2026} noted this shock, and the associated amplification of the magnetic field through the Weibel instability~\cite{Weibel1959}, as a natural next step, but did not compute the resulting emission.

The present paper attempts a first, necessarily approximate, estimate of that emission. Relativistic collisionless shocks lack a large-scale ordered field able to support efficient synchrotron emission or first-order Fermi acceleration; particle-in-cell (PIC) simulations and analytic arguments show that a two-stream (Weibel) instability in the shock transition layer generates small-scale, tangled magnetic fields on the scale of the plasma skin depth, saturating at $\epsB\sim10^{-5}$--$10^{-1}$ of equipartition~\cite{MedvedevLoeb1999,GruzinovWaxman1999,Spitkovsky2008,SironiSpitkovsky2011}. This is the same mechanism invoked to explain GRB afterglows~\cite{MeszarosRees1997,Waxman1997,SariPiranNarayan1998}, and it is the natural (if not the only conceivable) mechanism to invoke here as well.

Four simplifications keep the present model at the level of an order-of-magnitude estimate rather than a precision calculation, each revisited quantitatively below: (i) the isotropic kinetic energy that actually feeds the forward shock, $\eiso$, is only a fraction $R_k$ of the prompt-radiated energy computed by Villani~\cite{Villani2026}, the remainder having been radiated promptly or retained elsewhere (Sec.~\ref{sec:IC}); (ii) only the forward shock into the ambient electron-proton ISM is modeled, and the reverse shock into the pair-dominated ejecta is neglected, an approximation whose validity depends on the (poorly constrained) shell width (Sec.~\ref{sec:dynamics}); (iii) the SSC and $\gamma$-$\gamma$ opacity calculations use a one-zone, largely Thomson-regime treatment with an approximate Klein-Nishina correction (Sec.~\ref{sec:radiation}); and (iv) the detectability estimates use an illustrative flux/count threshold rather than a full instrument response (Sec.~\ref{sec:results}). Section~\ref{sec:IC} summarizes the initial conditions of Villani~\cite{Villani2026} and the $R_k$ parametrization; Section~\ref{sec:dynamics} presents the external-shock dynamics and the shell-width discussion; Section~\ref{sec:weibel} discusses Weibel-mediated field amplification and particle acceleration; Section~\ref{sec:radiation} gives the radiation, cooling, and opacity calculations; Section~\ref{sec:results} presents the numerical results; Sections~\ref{sec:discussion} and~\ref{sec:conclusions} discuss the uncertainties and conclude.

\section{Initial conditions from the black-to-white hole transition}
\label{sec:IC}

\subsection{Prompt radiated energy versus blast-wave kinetic energy}
\label{sec:energybudget}

The external shock is powered by whatever kinetic energy remains in the ejecta once the prompt photospheric flash has escaped, not by the prompt-radiated energy itself. Villani~\cite{Villani2026} self-consistently solves for the energy radiated in the prompt photospheric flash, $\egprompt$, finding $\egprompt\sim5\times10^{43}$--$3\times10^{44}$ erg depending on the assumed pion fraction and ejecta temperature. That calculation does not solve for the kinetic energy retained in the outflow once it becomes matter-dominated; that quantity, $\eiso$, which alone drives the external shock studied here, is not determined by the existing calculation.

We parametrize this unknown kinetic energy directly relative to the known prompt-radiated energy,
\begin{equation}
\eiso \equiv R_k\,\egprompt, \qquad R_k\in[10^{-3},1],
\label{eq:Rk}
\end{equation}
rather than introducing an independent, and currently unconstrained, total energy budget. Equivalently, defining the prompt radiative efficiency $\eta_\gamma\equiv\egprompt/(\egprompt+\eiso)$,
\begin{equation}
R_k = \frac{1-\eta_\gamma}{\eta_\gamma},
\label{eq:Rketa}
\end{equation}
so that $R_k\to0$ corresponds to a fully radiation-dominated outflow ($\eta_\gamma\to1$) and $R_k=1$ to equal partition between prompt-radiated and kinetic energy. We adopt $R_k=0.1$ as an illustrative fiducial value, comparable to the kinetic-to-prompt energy ratios inferred for classical GRBs~\cite{PanaitescuKumar2002,KumarZhang2015}, but Sec.~\ref{sec:results}B shows explicitly how the deceleration time, peak flux, and detectability depend on $R_k$ across the full range in Eq.~\eqref{eq:Rk}. A first-principles value of $R_k$ (equivalently $\eta_\gamma$) requires extending the photon-lepton kinetic solver of Villani~\cite{Villani2026} to track the kinetic energy retained in the matter-dominated outflow, which was not the focus of that calculation's radiative-transfer treatment, and is left for future work (Sec.~\ref{sec:discussion}).

\subsection{Coasting Lorentz factor and photospheric radius}

We take the coasting Lorentz factor $\Gamma_0$ of the matter-dominated outflow as a second free parameter, of order $10^2$--$10^3$, motivated by the low baryon loading found by Villani~\cite{Villani2026} ($n_b/n_{e^\pm}\sim1\%$) but not otherwise fixed by the present calculation; we adopt $\Gamma_0=100$ as fiducial. The photospheric radius $R_{\rm ph}$, at which the prompt flash of Ref.~\cite{Villani2026} is released, is set by the leptonic-scattering condition $\tau_{\rm es}(R_{\rm ph})\sim1$~\cite{Goodman1986,Paczynski1986} and is not quoted as an explicit number in that calculation. For a leptonic fireball of luminosity $L\sim\egprompt/T_{\rm eng}$ and Lorentz factor $\Gamma_0$, the standard compactness estimate gives
\begin{equation}
R_{\rm ph}\sim\frac{L\sigma_T}{8\pi\Gamma_0^3 m_e c^3}\sim10^{7}\text{--}10^{9}\ {\rm cm},
\label{eq:Rph}
\end{equation}
for engine timescales $T_{\rm eng}\sim10^{-3}$--$1$ s and the fiducial $\Gamma_0$; this is many orders of magnitude below the deceleration radius found in Sec.~\ref{sec:results} ($R_{\rm dec}\sim1$--$10$ AU $\sim10^{13}$--$10^{14}$ cm), which justifies treating the prompt photospheric emission and the external shock as well-separated in radius and time. We list this order-of-magnitude estimate, rather than a precise value, in Table~\ref{tab:fiducial}.

\begin{table}[t]
\caption{Fiducial parameter set adopted in Sec.~\ref{sec:results}, unless stated otherwise. $\eiso=R_k\egprompt$ (Sec.~\ref{sec:energybudget}); $R_{\rm ph}$ is an order-of-magnitude estimate (Eq.~\ref{eq:Rph}), not a value quoted by Villani~\cite{Villani2026}.}
\label{tab:fiducial}
\begin{ruledtabular}
\begin{tabular}{ll}
Parameter & Fiducial value \\
\hline
PBH mass, $M$ & $10^{26}$ g (lunar mass) \\
Prompt radiated energy, $\egprompt$ (Ref.~\cite{Villani2026}) & $10^{44}$ erg \\
Kinetic-to-prompt energy ratio, $R_k$ & $0.1$ (range $10^{-3}$--$1$) \\
Kinetic energy, $\eiso=R_k\egprompt$ & $10^{43}$ erg \\
Photospheric radius, $R_{\rm ph}$ (order of magnitude) & $10^{7}$--$10^{9}$ cm \\
Coasting Lorentz factor, $\Gamma_0$ & $100$ \\
ISM density, $\nism$ & $10^{-3},\,1,\,10^{2}~{\rm cm^{-3}}$ \\
Electron energy fraction, $\epse$ & $0.1$ \\
Electron acceleration fraction, $\xie$ & $1$ \\
Magnetic energy fraction, $\epsB$ & $10^{-3}$ (range $10^{-5}$--$10^{-1}$) \\
Electron spectral index, $p$ & $2.5$ \\
Source distance, $d$ & $1$ kpc (Galactic) \\
Redshift, $z$ & $0$ \\
\end{tabular}
\end{ruledtabular}
\end{table}

\section{External-shock dynamics}
\label{sec:dynamics}

\subsection{Lorentz-factor convention and dynamical prescription}

As the ejecta shell sweeps up circumstellar matter of constant density $\nism$, it decelerates. We use $\Gamma$ throughout to denote the Lorentz factor of the shocked fluid immediately behind the forward shock (the shock-front Lorentz factor differs by a fixed factor of $\sqrt{2}$ in the strong-shock limit, which does not affect any scaling used below, but which we flag here to fix the convention). We adopt the classical Blandford-McKee (BM) self-similar solution for a spherical, adiabatic, ultra-relativistic blast wave~\cite{BlandfordMcKee1976}, appropriate here because the near-isotropic geometry found by Villani~\cite{Villani2026} removes the need for jet collimation and because the radiative efficiency $\epse\lesssim0.1$ (Table~\ref{tab:fiducial}) keeps energy losses subdominant over the timescales of interest~\cite{PanaitescuKumar2002}. We emphasize that what follows is a piecewise coasting-plus-adiabatic-BM \emph{dynamical prescription} evaluated on a time grid, in the spirit of Sari, Piran \& Narayan~\cite{SariPiranNarayan1998}, rather than a full integration of the relativistic hydrodynamic equations.

The deceleration radius, where the swept-up rest-mass energy in the observer frame equals $\eiso/\Gamma_0^2$, is
\begin{equation}
R_{\rm dec} \simeq \left(\frac{3\eiso}{4\pi\nism m_p c^2\Gamma_0^2}\right)^{1/3},
\label{eq:Rdec}
\end{equation}
and the corresponding observer-frame deceleration time is
\begin{equation}
t_{\rm dec} \simeq (1+z)\,\frac{R_{\rm dec}}{2c\Gamma_0^2}.
\label{eq:tdec}
\end{equation}
For $t\le t_{\rm dec}$ the shell coasts at $\Gamma=\Gamma_0$. For $t>t_{\rm dec}$, energy conservation in the adiabatic BM self-similar solution, $\eiso=(17/16)\pi\nism m_p c^2\Gamma^2R^3$~\cite{BlandfordMcKee1976,Piran2005}, together with the observer-time relation $dt_{\rm obs}\simeq dR/(2c\Gamma^2)$, gives the closure-relation scalings~\cite{SariPiranNarayan1998,GaoEtal2013}
\begin{equation}
\Gamma(t) = \Gamma_0\left(\frac{t}{t_{\rm dec}}\right)^{-3/8}, \qquad
R(t) = R_{\rm dec}\left(\frac{t}{t_{\rm dec}}\right)^{1/4}.
\label{eq:GammaR}
\end{equation}
Because Eq.~\eqref{eq:GammaR} is only valid while the flow is ultra-relativistic, we restrict all light curves, radial profiles, and fluence integrals shown below to the window $t_{\rm dec}\le t\le t_{\rm valid}$, where $t_{\rm valid}$ is defined by $\Gamma(t_{\rm valid})=2.5$; the subsequent transition to a mildly, and eventually non-relativistic (Sedov-Taylor), blast wave is not modeled here and would require matching Eq.~\eqref{eq:GammaR} onto the corresponding non-relativistic self-similar solution. We caution that Eq.~\eqref{eq:GammaR} also assumes strict spherical symmetry and a negligible radiative loss fraction; both this and the relativistic-to-non-relativistic transition are revisited in Sec.~\ref{sec:discussion}.

\subsection{Forward shock composition versus pair-dominated ejecta}
\label{sec:composition}

Because the ejecta considered by Villani~\cite{Villani2026} are dominated by $e^\pm$ pairs and photons with only a $\sim1\%$ baryon fraction, it is important to be explicit about which plasma the electron-injection formulae of Sec.~\ref{sec:weibel} refer to. The forward shock propagates into the \emph{unshocked ambient ISM}, which we take to be ordinary electron-proton plasma; the standard injection formula $\gamma_m\propto m_p/m_e$ (Eq.~\ref{eq:gammam} below) reflects protons carrying most of the post-shock energy in this electron-proton medium, with a fraction $\epse$ transferred to the (here, ISM-supplied) electrons -- it is not a statement about the ejecta composition. The pair-dominated composition of the ejecta is instead directly relevant to the \emph{reverse} shock that forms in the ejecta itself (Sec.~\ref{sec:shellwidth}) and to the radiation-mediated shock physics of the prompt phase~\cite{Levinson2020}, neither of which is modeled quantitatively here. We further allow for the possibility that only a fraction $\xie\le1$ of the swept-up ISM electrons participate in the non-thermal acceleration process, with the rest remaining thermal (Sec.~\ref{sec:weibel}); we adopt $\xie=1$ as fiducial but note that $\epse/\xie$, not $\epse$ alone, controls the injection Lorentz factor.

\subsection{Shell width and the reverse shock}
\label{sec:shellwidth}

We model only the forward shock. This is a standard first approximation, but its validity depends on the shell-width regime~\cite{SariPiran1995,KobayashiSari2000}: writing the ejecta shell's lab-frame width as $\Delta\sim cT_{\rm eng}$, with $T_{\rm eng}$ the duration of the central engine (comparable to the prompt-emission duration), the shell is \emph{thin} if the reverse shock remains non-relativistic throughout, which requires $T_{\rm eng}\lesssim t_{\rm dec}$; otherwise the shell is \emph{thick} and the reverse shock becomes relativistic during crossing, altering the early forward-shock dynamics and potentially contributing a comparable or larger early-time optical/X-ray flash of its own~\cite{SariPiran1995,KobayashiSari2000}. For the fiducial parameters of Table~\ref{tab:fiducial}, $t_{\rm dec}\sim0.02$--$0.9$ s over the density range explored (Table~\ref{tab:peaks}); because the intrinsic duration of the black-to-white hole transition itself is not tightly constrained by the existing calculation~\cite{Villani2026}, $T_{\rm eng}$ could plausibly be comparable to, shorter than, or longer than $t_{\rm dec}$, placing the system in either regime depending on parameters we cannot yet fix. We therefore explicitly flag the results below as a \emph{forward-shock-only, lower-order estimate}: in the thick-shell case a relativistic reverse shock could contribute additional early-time emission not captured here, while in the thin-shell case the forward-shock treatment of Sec.~\ref{sec:radiation} should be adequate on its own. Resolving this requires an estimate of $T_{\rm eng}$ from the underlying dynamics of Ref.~\cite{Villani2026} that is not yet available.

\section{Magnetic-field generation and particle acceleration}
\label{sec:weibel}

\subsection{Weibel-mediated field amplification}

A relativistic collisionless shock propagating into an unmagnetized ISM cannot rely on flux-freezing of a pre-existing field to support synchrotron emission. Instead, the relative streaming between the incoming upstream plasma and the shock-heated downstream plasma is Weibel-unstable~\cite{Weibel1959}: current filaments grow on the scale of the local plasma skin depth and saturate into a small-scale, tangled magnetic field oriented predominantly transverse to the shock normal~\cite{MedvedevLoeb1999}. Modern particle-in-cell simulations of relativistic electron-ion shocks confirm that this mechanism both amplifies the field and mediates collisionless shock formation and particle acceleration in the absence of any large-scale field~\cite{Spitkovsky2008,GruzinovWaxman1999,SironiSpitkovsky2011}.

We parametrize the amplified comoving field strength phenomenologically as
\begin{equation}
\frac{B'^2}{8\pi} = \epsB\,e',
\label{eq:epsB}
\end{equation}
where $e'$ is the comoving internal energy density immediately behind the shock and $\epsB$ is the fraction of the shock energy channeled into magnetic turbulence. For a strong shock into the cold, uniform electron-proton ISM (Sec.~\ref{sec:composition}), the standard relativistic jump condition gives $e'\simeq4\Gamma^2\nism m_p c^2$~\cite{BlandfordMcKee1976,SariPiranNarayan1998}, so that
\begin{equation}
B'(t) \simeq \left(32\pi\,\epsB\,\Gamma(t)^2\,\nism\, m_p c^2\right)^{1/2}.
\label{eq:Bfield}
\end{equation}
We treat $\epsB$ as a free parameter in the theoretically and numerically motivated range $\epsB\sim10^{-5}$--$10^{-1}$~\cite{MedvedevLoeb1999}, constant with radius for the quantitative results presented here; a possible radial decline in the microphysical efficiency behind the shock~\cite{GaoEtal2013} is discussed qualitatively in Sec.~\ref{sec:discussion}.

\subsection{Electron acceleration}

We assume that a fraction $\xie$ of the swept-up ISM electrons are accelerated into a power law,
\begin{equation}
N(\gamma_e)\,d\gamma_e \propto \gamma_e^{-p}\,d\gamma_e, \qquad \gamma_e\ge\gamma_m,
\label{eq:Ndist}
\end{equation}
carrying a fraction $\epse$ of the shock energy. Sharing this energy among only the $\xie$ accelerated electrons (rather than all swept-up electrons) raises the minimum injection Lorentz factor by $1/\xie$ relative to the textbook $\xie=1$ case~\cite{SariPiranNarayan1998,GaoEtal2013}:
\begin{equation}
\gamma_m(t) = \frac{\epse}{\xie}\,\frac{p-2}{p-1}\,\frac{m_p}{m_e}\,\big(\Gamma(t)-1\big).
\label{eq:gammam}
\end{equation}
The factor $m_p/m_e$ enters because the forward shock propagates into the ordinary electron-proton ISM (Sec.~\ref{sec:composition}); it does not imply that the pair-dominated ejecta itself carries a proton-dominated internal energy. Radiative cooling defines a second characteristic Lorentz factor $\gamma_c$; because the SSC component computed in Sec.~\ref{sec:radiation} can carry a comparable or larger share of the cooling budget than synchrotron alone, we include the Compton parameter $Y$ (Sec.~\ref{sec:radiation}B) directly in the cooling rate,
\begin{equation}
\gamma_c(t) = \frac{6\pi m_e c\,(1+z)}{\sigma_T\,\Gamma(t)\,B'(t)^2\,t\,\big[1+Y(t)\big]},
\label{eq:gammac}
\end{equation}
where $\sigma_T$ is the Thomson cross section; $\gamma_c\to\gamma_{c,0}\equiv\gamma_c|_{Y=0}$ recovers the pure-synchrotron cooling rate used as a zeroth-order estimate in Sec.~\ref{sec:radiation}B. The relative ordering of $\gamma_m$ and $\gamma_c$ (equivalently of the corresponding frequencies $\nu_m$ and $\nu_c$, Sec.~\ref{sec:radiation}) determines whether the electron population is in the fast- or slow-cooling regime~\cite{SariPiranNarayan1998}.

\section{Synchrotron, inverse-Compton, and pair-opacity calculations}
\label{sec:radiation}

\subsection{Synchrotron spectrum}

The characteristic (observer-frame) synchrotron frequency emitted by an electron of Lorentz factor $\gamma_e$ in the comoving field $B'$, boosted by the bulk Lorentz factor $\Gamma$, is
\begin{equation}
\nu(\gamma_e,t) = \frac{3q_e}{4\pi m_e c}\,\frac{\Gamma(t)\,B'(t)\,\gamma_e^2}{1+z},
\label{eq:nuchar}
\end{equation}
which defines the injection frequency $\nu_m\equiv\nu(\gamma_m,t)$ and the cooling frequency $\nu_c\equiv\nu(\gamma_c,t)$, with $\gamma_c$ from Eq.~\eqref{eq:gammac}. The peak observed spectral flux density is~\cite{SariPiranNarayan1998,GaoEtal2013}
\begin{equation}
F_{\nu,\max}(t) = \frac{N_e(t)\,P'_{\nu,\max}(t)\,(1+z)}{4\pi d^2}, \qquad
P'_{\nu,\max} = \frac{m_e c^2\sigma_T}{3q_e}\,\Gamma(t)\,B'(t),
\label{eq:Fnumax}
\end{equation}
where $N_e(t)=\xie(4/3)\pi R(t)^3\nism$ is the number of accelerated electrons. The explicit factor of $\Gamma$ in $P'_{\nu,\max}$ (which boosts the comoving peak spectral power to the observer frame) is essential: together with $N_e\propto R^3\propto t^{3/4}$ and $B'\propto\Gamma\propto t^{-3/8}$, it makes $F_{\nu,\max}$ time-independent in the adiabatic ISM case, $F_{\nu,\max}\propto t^0$, a well-known closure relation of the standard afterglow model~\cite{SariPiranNarayan1998,Piran2005}. Above $\nu_a$ (Sec.~\ref{sec:selfabs}), the optically-thin spectrum is the usual smoothly-broken power law~\cite{SariPiranNarayan1998,GaoEtal2013},
\begin{equation}
F_\nu(t) = F_{\nu,\max}(t)\times
\begin{cases}
(\nu/\nu_1)^{1/3}, & \nu<\nu_1, \\[2pt]
(\nu/\nu_1)^{-s_1}, & \nu_1<\nu<\nu_2, \\[2pt]
\left(\dfrac{\nu_2}{\nu_1}\right)^{-s_1}\left(\dfrac{\nu}{\nu_2}\right)^{-p/2}, & \nu>\nu_2,
\end{cases}
\label{eq:Fnuseg}
\end{equation}
with $(\nu_1,\nu_2,s_1)=(\nu_c,\nu_m,1/2)$ in the fast-cooling regime ($\gamma_c<\gamma_m$) and $(\nu_1,\nu_2,s_1)=(\nu_m,\nu_c,(p-1)/2)$ in the slow-cooling regime ($\gamma_m<\gamma_c$).

\subsection{Synchrotron self-Compton emission, radiative efficiency, and the Compton parameter}
\label{sec:sscY}

Inverse-Compton scattering of the internal synchrotron photon field by the same shock-accelerated electrons -- SSC emission -- both radiates additional power and, through the associated IC cooling, shortens $\gamma_c$ [Eq.~\eqref{eq:gammac}]. We quantify its importance through the Compton parameter, computed self-consistently rather than assuming full radiative efficiency: let $\eta_{\rm rad}$ be the fraction of the injected electron energy that is radiated,
\begin{equation}
\eta_{\rm rad} = \begin{cases} 1, & \gamma_{c,0}<\gamma_m\ \text{(fast cooling)}, \\[2pt]
\left(\gamma_{c,0}/\gamma_m\right)^{2-p}, & \gamma_{c,0}>\gamma_m\ \text{(slow cooling)}, \end{cases}
\label{eq:etarad}
\end{equation}
evaluated using the zeroth-order (pure-synchrotron) cooling Lorentz factor $\gamma_{c,0}$ as a first-order estimate of the cooling regime~\cite{SariEsin2001}. The Thomson-regime Compton parameter is then
\begin{equation}
Y_{\rm Th} = \frac{1}{2}\left(-1+\sqrt{1+4\,\eta_{\rm rad}\,\epse/\epsB}\right),
\label{eq:Compton}
\end{equation}
which reduces to $Y_{\rm Th}\simeq\eta_{\rm rad}\epse/\epsB$ for $\eta_{\rm rad}\epse\ll\epsB$, and to $Y_{\rm Th}\simeq(\eta_{\rm rad}\epse/\epsB)^{1/2}$ in the opposite limit $\eta_{\rm rad}\epse\gg\epsB$~\cite{SariEsin2001}. Equation~\eqref{eq:Compton} feeds back into $\gamma_c$ through Eq.~\eqref{eq:gammac}, so $Y$, $\gamma_c$, and $\eta_{\rm rad}$ are evaluated self-consistently at first order (using $\gamma_{c,0}$ to fix the cooling regime, then correcting $\gamma_c$ by $1/(1+Y)$), rather than iterated to full convergence.

In the Thomson (single-scattering) approximation, the SSC component is a broken power law of the same shape as Eq.~\eqref{eq:Fnuseg}, with characteristic frequencies boosted by $\gamma_{m,c}^2$ and peak flux set by the shell's Thomson optical depth,
\begin{equation}
\nu_{m,c}^{\rm SSC,Th}\simeq\gamma_{m,c}^2\,\nu_{m,c}, \qquad
F_{\nu,\max}^{\rm SSC,Th}(t) \simeq \tau_T(t)\,F_{\nu,\max}(t), \qquad \tau_T=\nism\sigma_T R.
\label{eq:FSSC}
\end{equation}

\subsection{Klein-Nishina suppression}
\label{sec:KN}

Equation~\eqref{eq:FSSC} assumes Thomson-regime scattering, which requires the seed-photon energy in the electron's rest frame to be much less than $m_ec^2$. Defining the comoving seed-photon frequency $\nu'_m=\nu_m(1+z)/\Gamma$, the relevant Klein-Nishina parameter for an electron at the injection energy scattering its own synchrotron photons is
\begin{equation}
x_{\rm KN} \equiv \frac{4\gamma_m h\nu'_m}{m_ec^2}.
\label{eq:xKN}
\end{equation}
For the fiducial parameters of Table~\ref{tab:fiducial}, $x_{\rm KN}\sim4\times10^{-4}$--$0.1$ over the density range explored (Table~\ref{tab:peaks}): Klein-Nishina effects are sub-dominant but not always negligible, growing toward higher $\nism$ or $\epsB$. We include an order-of-magnitude correction, following the qualitative behavior established by the full analytic treatment of Nakar, Ando \& Sari~\cite{NakarAndoSari2009}: the effective Compton parameter is suppressed as $Y\simeq Y_{\rm Th}/(1+x_{\rm KN})$, the SSC peak flux is reduced by the same factor, and the up-scattered frequency is capped at the energy of the scattering electron itself, $\nu^{\rm SSC}\le\Gamma\gamma_{m,c}m_ec^2/[h(1+z)]$, since a single Compton scattering cannot transfer more energy than the electron possesses. This reproduces the correct order of magnitude and the qualitative hardening/softening trend of Ref.~\cite{NakarAndoSari2009} but is not a substitute for their full spectral solution, which would be needed for precision spectral fitting.

\subsection{Synchrotron self-absorption}
\label{sec:selfabs}

Below a self-absorption frequency $\nu_a$, the source becomes optically thick to its own synchrotron photons and the observed spectrum steepens from the optically-thin $F_\nu\propto\nu^{1/3}$ branch to a self-absorbed, Rayleigh-Jeans-like $F_\nu\propto\nu^{5/2}$ branch~\cite{RybickiLightman1979}. Following the brightness-temperature argument of Rybicki \& Lightman~\cite{RybickiLightman1979}, applied to a GRB-like shell as in Sari, Piran \& Narayan~\cite{SariPiranNarayan1998}: electrons radiating at comoving frequency $\nu'$ have Lorentz factor $\gamma(\nu')=(\nu'/\nu'_B)^{1/2}$, with $\nu'_B=q_eB'/(2\pi m_ec)$ the comoving cyclotron frequency, giving an effective brightness temperature $k_BT_{\rm eff}(\nu')\sim\gamma(\nu')m_ec^2/3$ and a self-absorbed flux $F_\nu^{\rm thick}(\nu)\propto\nu^{5/2}$ that we match numerically onto the optically-thin branch, Eq.~\eqref{eq:Fnuseg}, at $\nu=\nu_a$. We emphasize that this is an order-of-magnitude estimate of $\nu_a$, not a full radiative-transfer solution; the coefficient depends on order-unity choices in the brightness-temperature argument. For the density range explored (Table~\ref{tab:peaks}), $\nu_a$ ranges from $\sim1.6\times10^7$ Hz ($\nism=10^{-3}\,{\rm cm^{-3}}$) to $\sim1.3\times10^{10}$ Hz ($\nism=10^{2}\,{\rm cm^{-3}}$): at the highest densities considered, $\nu_a$ exceeds the fiducial radio band (1.4 GHz), so the naive optically-thin radio flux of Sec.~\ref{sec:results} is not a reliable prediction in that regime and the true radio flux is set instead by the self-absorbed branch, which we include explicitly in all light curves below.

\subsection{$\gamma$-$\gamma$ pair opacity}
\label{sec:gg}

High-energy synchrotron or SSC photons can pair-produce on the softer internal radiation field, $\gamma+\gamma\to e^-+e^+$, before escaping the source. The relevant optical depth is, in full generality,
\begin{equation}
\tau_{\gamma\gamma}(E_\gamma) = \int dl' \int d\epsilon' \int d\mu'\, n'(\epsilon',\mu')\,\sigma_{\gamma\gamma}(E'_\gamma,\epsilon',\mu')\,(1-\mu'),
\label{eq:tauggfull}
\end{equation}
where primes denote comoving-frame quantities, $l'$ is the comoving path length, $n'(\epsilon',\mu')$ is the angle-dependent target photon number density, and $\sigma_{\gamma\gamma}$ is the pair-production cross section~\cite{GouldSchreder1967}. Following the compactness argument used to derive lower limits on GRB Lorentz factors~\cite{LithwickSari2001}, we reduce Eq.~\eqref{eq:tauggfull} to an order-of-magnitude estimate by (i) evaluating the target photon field only at the pair-production threshold energy $\epsilon'_{\rm t}=(m_ec^2)^2/E'_\gamma$, (ii) approximating the angular integral over $(1-\mu')$ by an order-unity coefficient $\kappa\sim0.2$ appropriate for a quasi-isotropic comoving radiation field, and (iii) taking the comoving path length to be the shell's comoving radial extent, $l'\sim R/\Gamma$:
\begin{equation}
\tau_{\gamma\gamma}(E_\gamma,t) \simeq \kappa\,\sigma_T\,\frac{R(t)}{\Gamma(t)}\,n'_{\rm t}(\epsilon'_{\rm t},t),
\label{eq:taugg}
\end{equation}
with $n'_{\rm t}$ obtained self-consistently from the synchrotron-plus-SSC spectrum of Secs.~\ref{sec:radiation}A--\ref{sec:KN}. Equation~\eqref{eq:taugg} is explicitly an order-of-magnitude reduction of Eq.~\eqref{eq:tauggfull}, not a self-consistent radiative-transfer calculation; it is adequate to identify whether the source is qualitatively optically thin or thick to $\gamma$-$\gamma$ absorption at a given time and energy, which is the level at which we use it in Sec.~\ref{sec:results}.

\section{Results}
\label{sec:results}

We evaluate Eqs.~\eqref{eq:Rdec}--\eqref{eq:taugg} numerically for the fiducial parameters of Table~\ref{tab:fiducial} (now with $R_k=0.1$, i.e.\ $\eiso=10^{43}$ erg, rather than identifying $\eiso$ with the full prompt-radiated energy $\egprompt$) and for variations thereof, restricting all results to the ultra-relativistic validity window $\Gamma(t)\ge2.5$ (Sec.~\ref{sec:dynamics}A).

\subsection{Fiducial model}

\begin{figure}[t]
\centering
\includegraphics[width=\columnwidth]{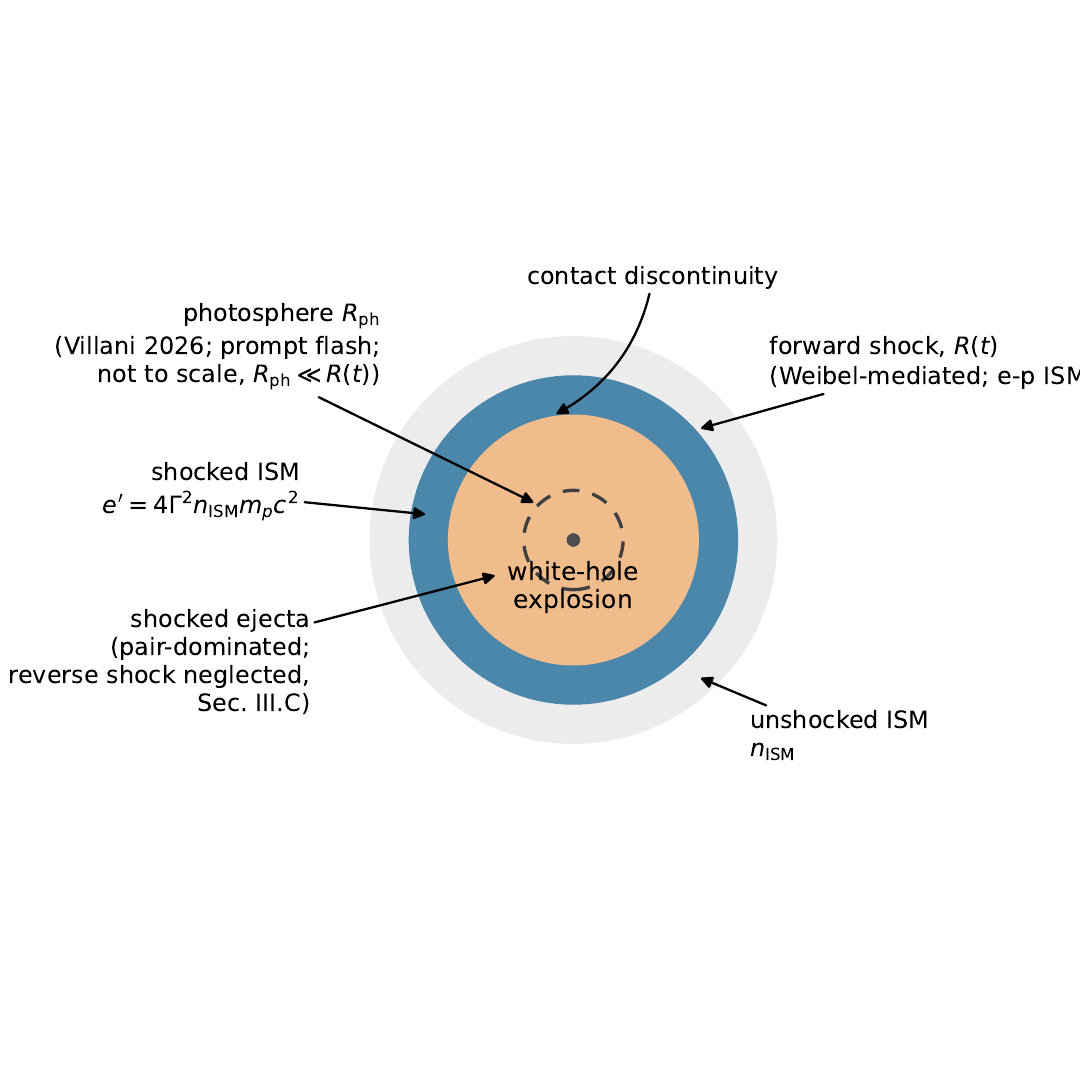}
\caption{Schematic of the source geometry (not to scale): the white-hole explosion is followed by the expanding photosphere of Villani~\cite{Villani2026} ($R_{\rm ph}\ll R(t)$, a much earlier and smaller-scale radius than the afterglow blast wave), a contact discontinuity, and a single Weibel-mediated shocked-ISM shell bounded by the contact discontinuity on the inside and the forward shock $R(t)$ on the outside, propagating into the unshocked ISM. The reverse shock into the pair-dominated ejecta (Sec.~\ref{sec:shellwidth}) is neglected in the present forward-shock-only estimate.}
\label{fig:schematic}
\end{figure}

\begin{figure}[t]
\centering
\includegraphics[width=\columnwidth]{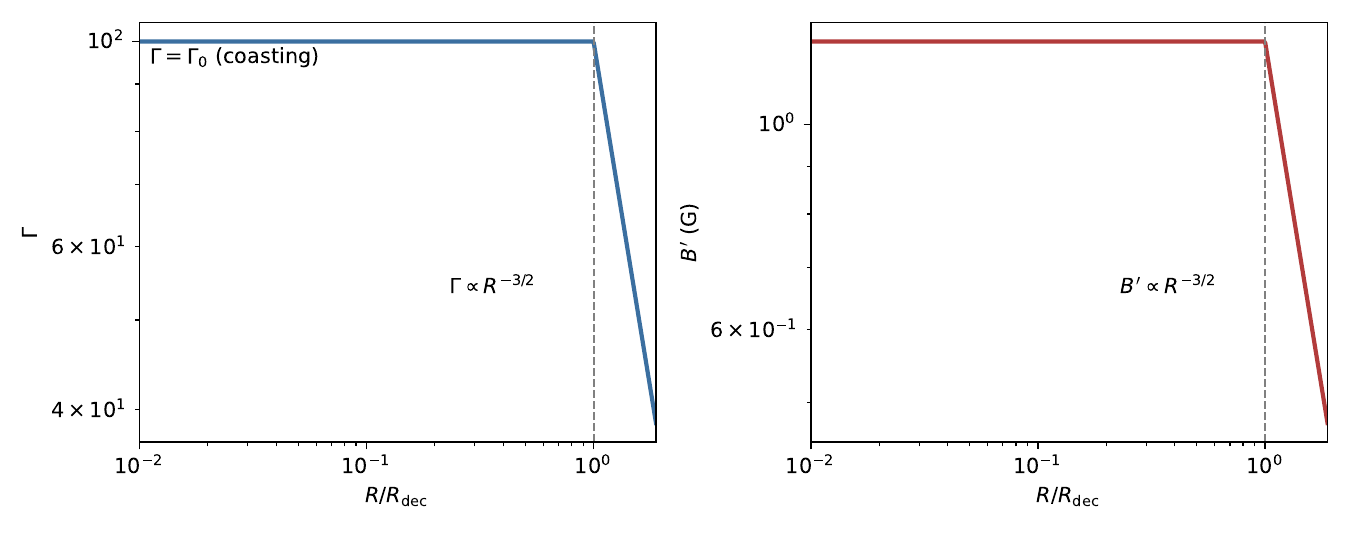}
\caption{(a) Bulk Lorentz factor $\Gamma$ and (b) Weibel-amplified comoving field $B'$ ($\epsB=10^{-3}$) against dimensionless radius $R/R_{\rm dec}$, for the fiducial model. Dotted lines show the asymptotic $R^{-3/2}$ scaling; curves are shown only over the ultra-relativistic window $\Gamma\ge2.5$ (Sec.~\ref{sec:dynamics}A).}
\label{fig:GammaB}
\end{figure}

Figure~\ref{fig:schematic} sketches the emission geometry, including the contact discontinuity and the (neglected) reverse shock discussed in Sec.~\ref{sec:shellwidth}. For the fiducial parameters, deceleration occurs at
\begin{equation}
R_{\rm dec}\simeq3.6~{\rm AU}, \qquad t_{\rm dec}\simeq0.090~{\rm s}\qquad(\nism=1~{\rm cm^{-3}}).
\label{eq:fiducial_dec}
\end{equation}
Figure~\ref{fig:GammaB} shows $\Gamma$ and $B'$ against the dimensionless radius $R/R_{\rm dec}$: $\Gamma$ is constant during the coasting phase and falls as $\Gamma\propto R^{-3/2}$ thereafter (dotted guide line), while $B'$ follows the same scaling through Eq.~\eqref{eq:Bfield}.

\begin{figure}[t]
\centering
\includegraphics[width=\columnwidth]{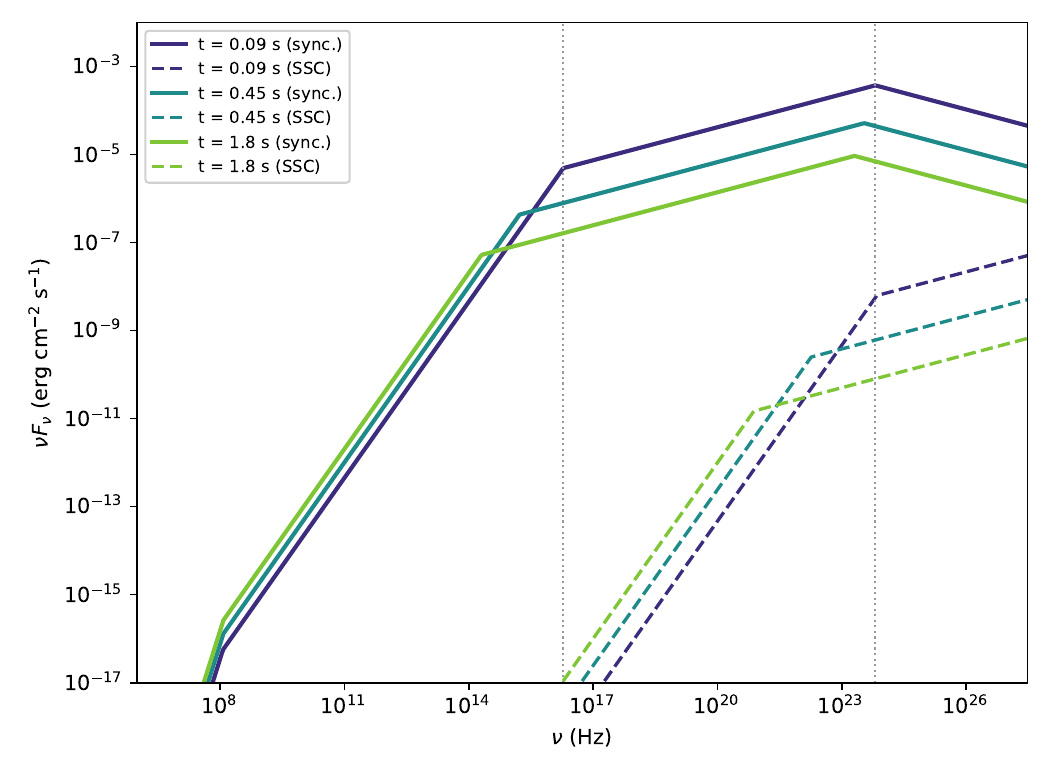}
\caption{Synchrotron (solid) and SSC (dashed) spectral energy distributions at three observer times, in the Thomson approximation with the Klein-Nishina correction of Sec.~\ref{sec:KN}; $\gamma$-$\gamma$ attenuation is not applied to the curves shown (Sec.~\ref{sec:gg} gives $\tau_{\gamma\gamma}$ separately). Vertical dotted lines mark $\nu_a$, $\nu_1$, and $\nu_2$ at $t=t_{\rm dec}$.}
\label{fig:spectra}
\end{figure}

Figure~\ref{fig:spectra} shows the synchrotron (solid) and SSC (dashed) spectral energy distributions at three times, with the self-absorption break $\nu_a$ and the two spectral breaks $\nu_1<\nu_2$ marked at the reference time $t=t_{\rm dec}$. The SSC curves include the order-of-magnitude Klein-Nishina suppression of Sec.~\ref{sec:KN}; neither curve includes $\gamma$-$\gamma$ attenuation, which is computed separately as $\tau_{\gamma\gamma}(E_\gamma,t)$ (Sec.~\ref{sec:gg}) and is not folded into the displayed spectra. Over the parameter range explored the electron population is in the slow-cooling regime ($\gamma_{c,0}\gg\gamma_m$, so $\eta_{\rm rad}\ll1$) at all times shown, so $(\nu_1,\nu_2)=(\nu_m,\nu_c)$.

\subsection{Dependence on ambient density and on the kinetic efficiency $R_k$}

\begin{figure}[t]
\centering
\includegraphics[width=\columnwidth]{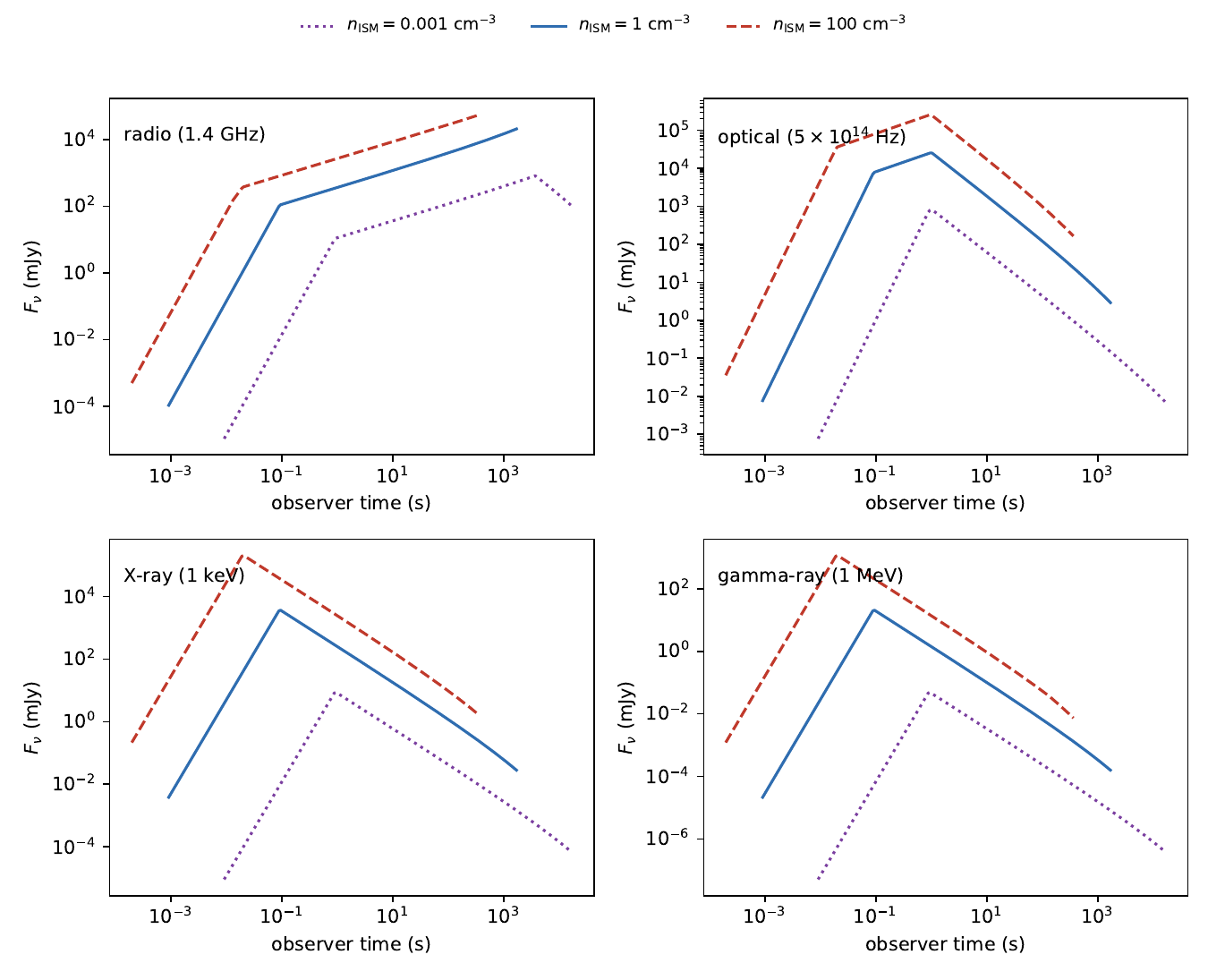}
\caption{Multiwavelength light curves, one panel per band, for three ambient densities, including synchrotron self-absorption (Sec.~\ref{sec:selfabs}) and truncated at $\Gamma=2.5$ (Sec.~\ref{sec:dynamics}A).}
\label{fig:lightcurves}
\end{figure}

Figure~\ref{fig:lightcurves} shows multiwavelength light curves, now with synchrotron self-absorption included (Sec.~\ref{sec:selfabs}) and split into one panel per band for clarity, for $\nism=10^{-3},\,1,\,10^{2}~{\rm cm^{-3}}$. Table~\ref{tab:peaks} lists $t_{\rm dec}$, $F_{\nu,\max}$, $\nu_a$, and $x_{\rm KN}$ for each density. As anticipated in Sec.~\ref{sec:selfabs}, at $\nism=10^2\,{\rm cm^{-3}}$ the self-absorption frequency $\nu_a\sim1.3\times10^{10}$ Hz lies \emph{above} the fiducial radio band, so the radio curve in that case is entirely on the self-absorbed branch and is substantially fainter than a naive optically-thin extrapolation would suggest; at $\nism=10^{-3}\,{\rm cm^{-3}}$, by contrast, $\nu_a\ll1.4$ GHz and the radio band is optically thin throughout.

\begin{table}[t]
\caption{Deceleration time, peak synchrotron flux density, self-absorption frequency, and Klein-Nishina parameter (Sec.~\ref{sec:KN}) at $t=2t_{\rm dec}$, for the three fiducial densities of Fig.~\ref{fig:lightcurves} ($d=1$ kpc, $R_k=0.1$). In all three cases the electrons are slow-cooling ($\gamma_{c,0}/\gamma_m\approx6\times10^{2}$--$1\times10^{4}$ at $t=2t_{\rm dec}$, so $\eta_{\rm rad}\ll1$).}
\label{tab:peaks}
\begin{ruledtabular}
\begin{tabular}{ccccc}
$\nism$ (cm$^{-3}$) & $t_{\rm dec}$ (s) & $F_{\nu,\max}$ (mJy) & $\nu_a$ (Hz) & $x_{\rm KN}$ \\
\hline
$10^{-3}$ & $0.90$ & $8.2\times10^{2}$ & $1.6\times10^{7}$ & $4\times10^{-4}$ \\
$1$ & $0.090$ & $2.6\times10^{4}$ & $8.7\times10^{8}$ & $1.3\times10^{-2}$ \\
$10^{2}$ & $0.019$ & $2.6\times10^{5}$ & $1.3\times10^{10}$ & $0.13$ \\
\end{tabular}
\end{ruledtabular}
\end{table}

\begin{figure}[t]
\centering
\includegraphics[width=\columnwidth]{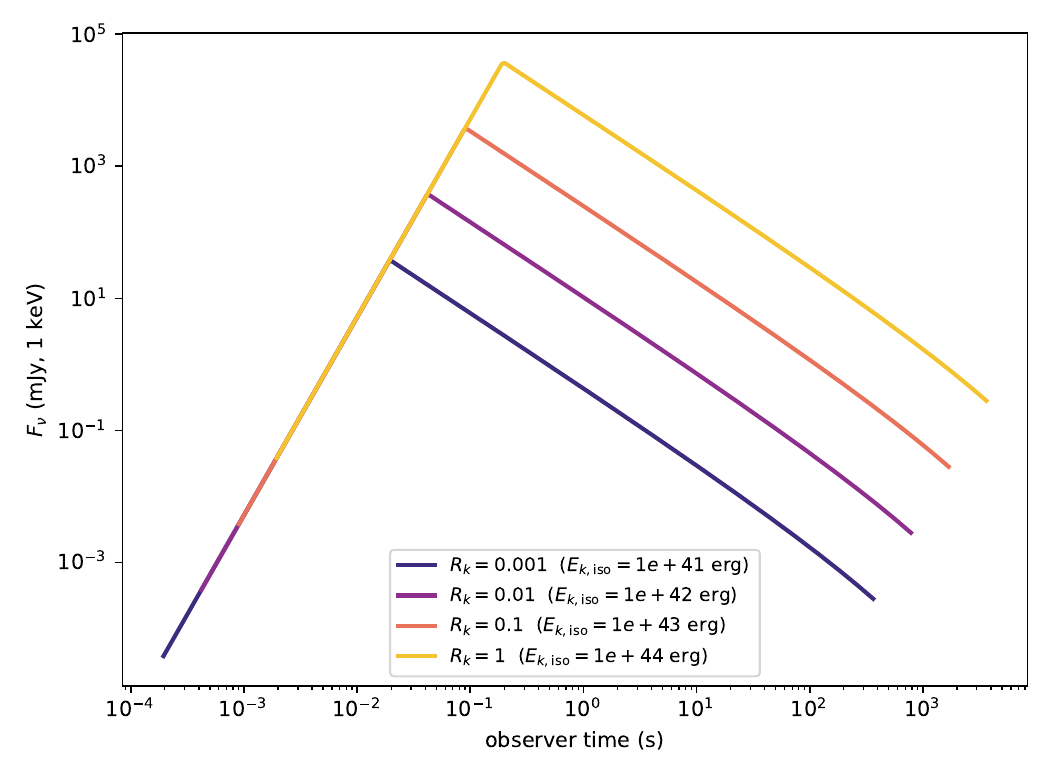}
\caption{X-ray (1 keV) light curve as a function of the kinetic efficiency $R_k=\eiso/\egprompt$, for $\egprompt=10^{44}$ erg and otherwise fiducial parameters.}
\label{fig:fkscan}
\end{figure}

Figure~\ref{fig:fkscan} shows the effect of the kinetic-to-prompt energy ratio $R_k$ directly: because $R_{\rm dec}\propto\eiso^{1/3}\propto R_k^{1/3}$ and $F_{\nu,\max}\propto\eiso$ [through $N_e$ at fixed $\Gamma_0,\nism$; Eq.~\eqref{eq:Fnumax}], the X-ray light curve shifts to shorter timescales and lower flux as $R_k$ decreases from $1$ to $10^{-3}$, spanning nearly three orders of magnitude in peak flux over the range considered. This directly addresses the concern that $\eiso$ should not be fixed to the full prompt-radiated energy: the detectability conclusions of Sec.~\ref{sec:results}E depend sensitively on $R_k$, which is not yet independently constrained.

\subsection{Dependence on magnetic-field efficiency}

Because $B'\propto\epsB^{1/2}$ [Eq.~\eqref{eq:Bfield}], both $\nu_{m,c}$ and $F_{\nu,\max}$ depend on $B'$ [Eqs.~\eqref{eq:nuchar}--\eqref{eq:Fnumax}]. Across the theoretically motivated range $\epsB\sim10^{-5}$--$10^{-1}$~\cite{MedvedevLoeb1999}, $F_{\nu,\max}\propto\epsB^{1/2}$ changes by two orders of magnitude, while $\nu_c$ is more sensitive still, since increasing $\epsB$ also increases $x_{\rm KN}$ [Eq.~\eqref{eq:xKN}, through $\nu_m\propto B'$] and hence modifies the Klein-Nishina suppression of $Y$. For the fiducial $\Gamma_0=100$, the system remains slow-cooling ($\gamma_{c,0}\gg\gamma_m$) throughout the entire theoretically motivated range $\epsB\sim10^{-5}$--$10^{-1}$ and the full $\nism$ range explored; a transition to fast cooling requires both a higher $\Gamma_0\sim10^3$ and $\epsB$ near the top of this range simultaneously, and is not reached by any of the fiducial cases shown in Figs.~\ref{fig:lightcurves}--\ref{fig:fkscan}.

\subsection{Multiwavelength light curves and the reverse-shock caveat}

The light curves in Fig.~\ref{fig:lightcurves} show a generic pattern: an approximately flat phase before the relevant characteristic frequency crosses the observing band, followed by a power-law decay once $\nu_m$ (or $\nu_c$) drops below it. For the fiducial parameters, the forward-shock afterglow evolves on timescales of order $0.02$--$0.9$ s to a few $\times10^{2}$--$10^4$ s, faster than the day-to-month timescales of classical long-GRB afterglows because of the much lower isotropic energy of the PBH transition relative to $\eiso\sim10^{51-54}$ erg typical of GRBs~\cite{Piran2005,KumarZhang2015}. As discussed in Sec.~\ref{sec:shellwidth}, if the (currently unconstrained) engine timescale $T_{\rm eng}$ exceeds $t_{\rm dec}$, a relativistic reverse shock could add early-time emission not included in Fig.~\ref{fig:lightcurves}; we regard the forward-shock light curves shown here as a lower bound on the early-time flux in that case.

\subsection{Detectability: an illustrative count-based criterion}

\begin{figure}[t]
\centering
\includegraphics[width=\columnwidth]{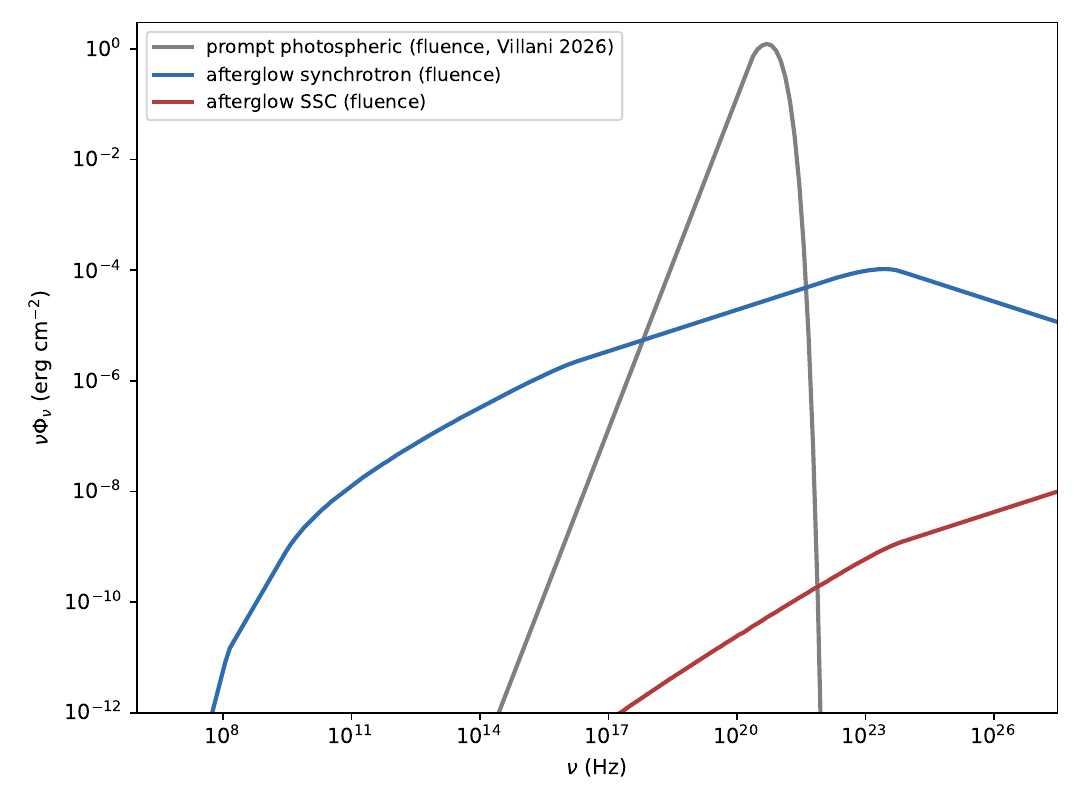}
\caption{Prompt photospheric energy fluence~\cite{Villani2026} compared with the afterglow synchrotron and SSC fluence integrated over $t_{\rm dec}\le t\le t_{\rm valid}$, both in absolute units (erg cm$^{-2}$), for $R_k=0.1$.}
\label{fig:compare}
\end{figure}

Figure~\ref{fig:compare} compares the prompt photospheric emission of Villani~\cite{Villani2026} with the afterglow synchrotron and SSC emission computed here, using the absolute energy fluence (erg cm$^{-2}$) in both cases -- the prompt fluence from the actual radiated energy $\egprompt$ distributed over the spectral shape of Ref.~\cite{Villani2026}, and the afterglow fluence from the synchrotron and SSC flux integrated over the ultra-relativistic window $t_{\rm dec}\le t\le t_{\rm valid}$, with $t_{\rm valid}$ defined by $\Gamma(t_{\rm valid})=2.5$ (Sec.~\ref{sec:dynamics}A) -- rather than an arbitrarily normalized snapshot. This comparison excludes the pre-deceleration coasting contribution ($t<t_{\rm dec}$) and should therefore be read as a conservative estimate of the forward-shock fluence. For the fiducial parameters, the total afterglow fluence is several orders of magnitude below the prompt fluence, reflecting the product of microphysical efficiencies $\epse\epsB^{1/2}\ll1$; the SSC component is smaller still once the Klein-Nishina and Thomson-depth suppressions of Sec.~\ref{sec:KN} are included.

\begin{figure}[t]
\centering
\includegraphics[width=\columnwidth]{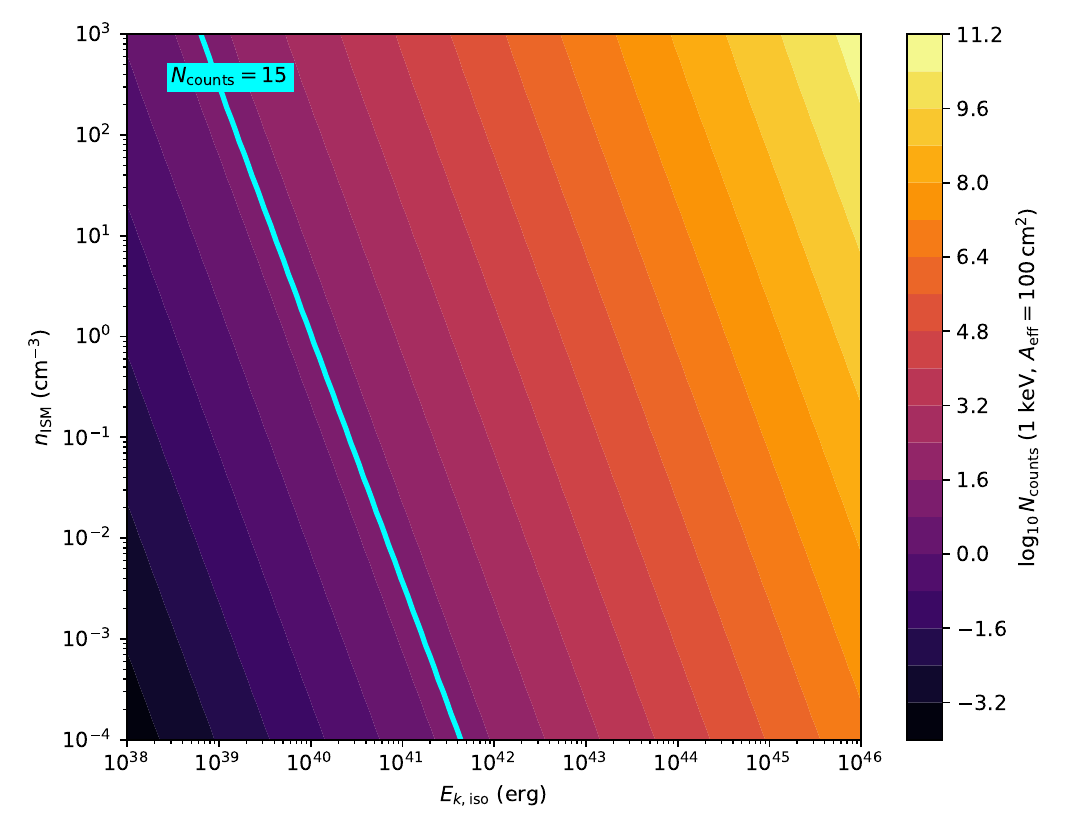}
\caption{Illustrative photon-count estimate [Eq.~\eqref{eq:counts}] at 1 keV, including Galactic photoelectric absorption, as a function of $\eiso$ and $\nism$, for $d=1$ kpc and $A_{\rm eff}=100\ {\rm cm^2}$. The cyan contour marks $N_{\rm counts}=15$; this is not a full instrument-trigger calculation (Sec.~\ref{sec:results}E).}
\label{fig:detectability}
\end{figure}

Figure~\ref{fig:detectability} replaces the bare flux-threshold comparison of a peak-flux map with an illustrative photon-count estimate,
\begin{equation}
N_{\rm counts} \sim \frac{A_{\rm eff}}{h}\int_{t_{\rm dec}}^{t_{\rm valid}} F_\nu(t)\,dt\;e^{-N_H\sigma_{\rm ISM}(E)},
\label{eq:counts}
\end{equation}
evaluated at 1 keV for a representative focusing-telescope effective area $A_{\rm eff}=100\ {\rm cm^2}$ -- illustrative only, and not intended to represent the triggering performance of a specific mission -- including an approximate Galactic photoelectric absorption factor using the solar-abundance Morrison \& McCammon~\cite{MorrisonMcCammon1983} cross section $\sigma_{\rm ISM}(E)\approx2\times10^{-22}(E/{\rm keV})^{-8/3}\ {\rm cm^2}$ and a fiducial line-of-sight column $N_H=3\times10^{21}\ {\rm cm^{-2}}$, representative of $d\sim1$ kpc but not tied to any specific line of sight; this Galactic-average quantity is independent of the local circumstellar density $\nism$ that controls the shock physics, and Fig.~\ref{fig:detectability} should accordingly be read as a fiducial-column illustration rather than a universal detectability map. At 1 keV this absorption removes $\sim45\%$ of the flux for the fiducial column. $N_{\rm counts}$ is compared to an illustrative minimum of $N_{\rm min}=15$ counts, appropriate for a marginal detection in a low-background, short-exposure search. Figure~\ref{fig:detectability} is an illustrative count-rate estimate, not a full instrument-trigger calculation, which would additionally require the detector background rate, point-spread function, and search-cadence-dependent trials factor.

For $\eiso\gtrsim10^{41}$--$10^{42}$ erg (i.e.\ $R_k\gtrsim10^{-3}$--$10^{-2}$ at fixed $\egprompt=10^{44}$ erg) and $\nism\gtrsim10^{-2}\ {\rm cm^{-3}}$, $N_{\rm counts}$ exceeds our illustrative threshold, but the sub-second-to-minute duration means detection would require sub-second-cadence monitoring or serendipitous coverage rather than the day-scale follow-up used for GRB afterglows, and the true detectability additionally depends on the instrument-specific background and trigger algorithm not modeled here.

\section{Discussion}
\label{sec:discussion}

Several assumptions deserve to be highlighted as genuine sources of uncertainty rather than as free normalization choices.

\emph{Kinetic versus prompt energy.} The single most consequential uncertainty identified in this revision is the split between $\egprompt$ and $\eiso$ (Sec.~\ref{sec:energybudget}): we have parametrized it through $R_k$ and shown its effect explicitly (Fig.~\ref{fig:fkscan}), but a first-principles value of $R_k$ (equivalently $\eta_\gamma$) requires extending the photon-lepton kinetic solver of Villani~\cite{Villani2026} to track the kinetic energy retained in the matter-dominated outflow, which was not the focus of that calculation's radiative-transfer treatment.

\emph{Shell width and the reverse shock.} We have modeled only the forward shock (Sec.~\ref{sec:shellwidth}); whether this is adequate depends on the ratio $T_{\rm eng}/t_{\rm dec}$, which we cannot yet fix from first principles. A thick-shell, relativistic reverse shock would add an additional early-time, and possibly brighter, transient component not captured here~\cite{SariPiran1995,KobayashiSari2000}.

\emph{Weibel microphysics and $\epsB$ evolution.} We have treated $\epsB$ as constant in radius, but PIC simulations and analytic arguments suggest the microphysical efficiency of Weibel-generated turbulence may decline behind the shock front~\cite{GaoEtal2013,MedvedevLoeb1999,SironiSpitkovsky2011}. Because $F_{\nu,\max}\propto\epsB^{1/2}$ and $\nu_c$ depends sensitively on $\epsB$ through both $B'$ and $Y$ (Sec.~\ref{sec:results}C), a declining $\epsB(R)$ would soften the light-curve decay without changing the overall order of magnitude of the peak flux.

\emph{One-zone, Thomson-regime SSC and $\gamma$-$\gamma$ opacity.} The SSC and pair-opacity calculations of Secs.~\ref{sec:sscY}--\ref{sec:gg} use a one-zone approximation and an approximate (not fully spectrally resolved) Klein-Nishina correction~\cite{NakarAndoSari2009}; a precision treatment of the GeV-band detectability discussed qualitatively in Sec.~\ref{sec:results}E would require the full KN-corrected spectral solution.

\emph{Geometry and the jet-break signature.} The assumption of strict spherical symmetry, motivated by the absence of a large-scale magnetic field around an isolated PBH~\cite{Villani2026}, removes the achromatic jet break characteristic of collimated GRB outflows~\cite{SariPiranHalpern1999}. The \emph{absence} of a jet break is consistent with, but not a unique signature of, a quasi-spherical outflow: several other classes of quasi-spherical transients share this feature, so it should be read as one discriminant among several rather than a decisive one.

\emph{Interstellar absorption.} Soft X-ray and EUV photons from a Galactic source are subject to significant photoelectric absorption by the intervening ISM~\cite{MorrisonMcCammon1983}; we included an approximate correction in Fig.~\ref{fig:detectability} ($\sim45\%$ absorbed at 1 keV for our fiducial column), but a full treatment would require the source's actual Galactic coordinates and the corresponding $N_H$ along that specific line of sight, which are not defined for the generic source considered here.

\emph{Redshift, rate, and selection effects.} We have focused on a Galactic ($z=0$) source; Villani~\cite{Villani2026} additionally constrains the cosmological density of PBHs that have undergone the transition, $\Omega_{\rm PBH}\sim10^{-8}$, using diffuse MeV background data. Extending the present afterglow calculation to a cosmological population is left for future work.

The energetic hierarchy underlying the qualitative picture is worth restating: because $\eiso$ is many orders of magnitude below the isotropic kinetic energies of long GRBs even at $R_k=1$, the non-thermal afterglow -- whatever its precise normalization once $R_k$, $\xie$, and the shell-width regime are pinned down -- is expected to evolve on sub-minute-to-hour timescales (Table~\ref{tab:peaks}) rather than the day-to-month timescales of GRB afterglows. This feature -- a fast, comparatively faint, roughly isotropic non-thermal transient following (or overlapping with) a hard-spectrum, sub-second-to-second MeV flash -- is the most robust qualitative discriminant identified so far, although the quantitative flux and duration predicted here remain order-of-magnitude estimates pending the improvements outlined above.

\section{Conclusions}
\label{sec:conclusions}

We have extended the photospheric model of Villani~\cite{Villani2026} with a first, explicitly approximate, estimate of the non-thermal emission produced when the relativistic, near-isotropic kinetic energy left over from a black-to-white hole transition collides with the interstellar medium. Modeling the resulting forward shock with the Blandford-McKee self-similar solution~\cite{BlandfordMcKee1976} and a Weibel-mediated magnetic field~\cite{Weibel1959,MedvedevLoeb1999,SironiSpitkovsky2011}, we estimated the synchrotron and SSC spectra, light curves, and $\gamma$-$\gamma$ pair opacity as functions of the ambient density, the kinetic-to-prompt energy ratio $R_k$, and the microphysical parameters $\epse$, $\epsB$, $\xie$, and $p$, now including a first-order IC-cooling correction, an approximate Klein-Nishina suppression, and synchrotron self-absorption.

We find that: (i) a non-thermal afterglow can exist and, for $R_k\gtrsim10^{-3}$--$10^{-2}$ and $\nism\gtrsim10^{-2}\,{\rm cm^{-3}}$, an illustrative photon-count estimate exceeds a nominal detection threshold out to $\sim1$ kpc (Sec.~\ref{sec:results}E), though this is sensitive to $R_k$, which is not yet constrained by the existing calculation~\cite{Villani2026}; (ii) its intrinsic spectral energy distribution peaks in the soft X-ray/EUV band near deceleration and evolves toward lower frequencies within seconds to minutes, with Klein-Nishina corrections that remain modest ($x_{\rm KN}\lesssim0.1$) but non-negligible over the density range explored -- Galactic photoelectric absorption can strongly suppress the observable EUV flux, however, so the practical discovery band for a nearby Galactic event shifts toward soft X-rays; (iii) synchrotron self-absorption can suppress the radio flux by orders of magnitude at high ambient density, so radio predictions should not be taken as reliable without this correction; (iv) for the adopted energy, density, and Lorentz-factor range, its duration is several orders of magnitude shorter than for classical long-GRB afterglows; and (v) the absence of jet collimation removes the achromatic jet break otherwise expected in a beamed outflow, though this alone is not a unique signature. The largest remaining uncertainties -- the kinetic-to-prompt energy ratio ($R_k$), the shell-width regime and the possible role of a reverse shock, and a fully spectrally-resolved Klein-Nishina/SSC treatment -- are identified explicitly as targets for future work, together with a cosmological population study building on the rate constraints of Villani~\cite{Villani2026}.

\begin{acknowledgments}
The author is grateful to Professor Tran Huu Phat for stimulating discussions and valuable comments.
\end{acknowledgments}
\section*{Funding Declaration}
The author declares that no funding, grants, or other support was received during the preparation of this manuscript.

\end{document}